# Muon Spin Relaxation Study of Spin Dynamics on a Kitaev honeycomb material $H_3LiIr_2O_6$

Yan-Xing Yang[1], Cheng-Yu Jiang[1], Liang-Long Huang[2], Zi-Hao Zhu[1], Chang-Sheng Chen[1], Qiong Wu[1], Zhao-Feng Ding[1], Cheng Tan[1], Kai-Wen Chen[1], Pabi K. Biswas[3], Adrian D. Hillier[3], You-Guo Shi[4, 5], Cai Liu[2, 6], Le Wang[2, 6], Fei Ye[2, 6], Jia-Wei Mei[2, 6], and Lei Shu[1, 7]

[1]State Key Laboratory of Surface Physics, Department of Physics, Fudan University, Shanghai 200438, China

[2]Department of Physics, Southern University of Science and Technology, Shenzhen 518055, China

[3]ISIS Facility, STFC Rutherford Appleton Laboratory, Chilton, Didcot, Oxfordshire, OX110QX, United Kingdom

[4]Institute of Physics, Chinese Academy of Sciences, Beijing 100190, China

[5]School of Physical Sciences, University of Chinese Academy of Sciences, Beijing 100190, China

[6]Shenzhen Key Laboratory of Quantum Science and Engineering, Shenzhen 518055, PR China.

[7]Collaborative Innovation Center of Advanced Microstructures, Nanjing 210093, China

## ABSTRACT

The vacancy effect in quantum spin liquid (QSL) has been extensively studied. A finite density of random vacancies in the Kitaev model can lead to a pileup of low-energy density of states (DOS), which is generally experimentally determined by a scaling behavior of thermodynamic or magnetization quantities. Here, we report detailed muon spin relaxation ($\mu$SR) results of $H_3LiIr_2O_6$, a Kitaev QSL candidate with vacancies. The absence of magnetic order is confirmed down to 80 mK, and the spin fluctuations are found to be persistent at low temperatures. Intriguingly, the time-field scaling law of longitudinal-field (LF)-$\mu$SR polarization is observed down to 0.1 K. This indicates a dynamical scaling, whose critical exponent 0.46 is excellently consistent with the scaling behavior of specific heat and magnetization data. All the observations point to the finite DOS with the form $N(E) \sim E^{-\nu}$, which is expected for the Kitaev QSL in the presence of vacancies. Our $\mu$SR study provides a dynamical fingerprint of the power-law low-energy DOS, and introduces a crucial new insight into the vacancy effect in QSL.

## INTRODUCTION

Quantum spin liquid (QSL) is a highly entangled quantum state in the spin system where the spin degree of freedom does not freeze even at zero temperature, but highly entangles with each other [1-8]. The delicate many-body entanglement in QSL is a crucial ingredient for the mechanism of high-temperature superconductivity [2] and the implementation of topological quantum computation [9]. The exactly solvable Kitaev honeycomb spin model [3] establishes the very existence of QSL in a simple spin interacting system, and the materialization of the Kitaev QSL in the experiments has been initialized currently [10-12]. With the help of the intertwining among magnetism, spin-orbital coupling, and crystal field, $Ir^{4+}$ oxides and a $Ru^{3+}$ chloride with a $d^5$ electronic configuration are promising to materialize the Kitaev model [13-16]. However, due to the non-Kitaev terms, these Kitaev compounds (e.g., $\alpha$-$Li_2IrO_3$ and $\alpha$-$RuCl_3$) usually develop long-range magnetic order at low temperatures [13,17,18]. Although there are several experimental signatures of fractionalized high-energy spin excitations [19-21], magnetic order precludes the access to the low-energy and low-temperature properties of QSL.

Chemical modification of $\alpha$-$Li_2IrO_3$ leads to the second generation of two-dimensional iridates $H_3LiIr_2O_6$ with a honeycomb lattice (see Fig. 1a) [22-24]. Thermodynamic and nuclear magnetic resonance (NMR) measurements did not detect any magnetic order down to 50 mK, indicating a possible QSL ground state in $H_3LiIr_2O_6$ [23]. Magnetic Raman spectroscopic study on single-crystal $H_3LiIr_2O_6$ samples reveals a dome-shaped magnetic Raman continuum [24] related to the high-energy fractionalized spin excitations (≥ 2 meV) [25]. However, $H_3LiIr_2O_6$ has various forms of stacking faults [22-24], the randomness of intercalated H positions [24,26], and non-magnetic $Ir^{3+}$ defects with a lower-oxidation-state configuration $3d^6$ [24]. The first-generation iridate $\alpha$-$Li_2IrO_3$ already manifests stacking faults along the $c$-axis [20,27]. The chemical modification further increases the stacking faults in the second-generation iridate $H_3LiIr_2O_6$ [22-24,28], indicated by the broad and weak powder X-ray diffraction (PXRD) peaks shown in Fig. 1b. The vacancies mentioned in this article include true vacancies due to the presence of non-magnetic impurities and quasivacancies originated form bond randomness. This description follows the theoretical work on the random-vacancies scenario in Kitaev QSL [29]. The vacancies can suppress the long-range magnetic order and release the QSL properties covered by the magnetic order, leading to new phenomena as a cooperative manifestation of disordered topological condensed matter systems [29-36]. In $H_3LiIr_2O_6$, at odds with the thermodynamics of pure Kitaev QSL, the abundant low-energy DOS is observed in the heat capacity where the specific heat coefficient displays a low-temperature divergence of $C/T \sim T^{-0.5}$ below 1 K [23], which is a characteristic of Kitaev QSL with vacancies [23,29,36]. To further investigate the spin dynamics and characterize the exotic spin correlation related to the vacancy-induced low-energy DOS in $H_3LiIr_2O_6$, we carried out $\mu$SR measurements.

This work reports detailed $\mu$SR results on $H_3LiIr_2O_6$. Consistent with the previous studies [23], our heat capacity measurements do not detect any magnetic order down to 0.1 K, and the scaling behavior is observed in both the magnetization and specific heat. The ZF-$\mu$SR confirms no sign of magnetic ordering, LF-$\mu$SR indicates dynamic spin fluctuations down to 80 mK. To establish the correlation between the low-energy spin fluctuations and the vacancy-induced DOS, magnetic field dependence of spin dynamics is studied and the time-field scaling behavior of LF-$\mu$SR polarization, $A(t) \sim t/B^{\alpha}$ ($\alpha = 0.46$) is revealed at low applied magnetic fields. This implies the scaling of local spin dynamics $q(t) = \langle S_i(t) \cdot S_i(0) \rangle \sim t^{\alpha-1} \exp(-\lambda t)$. The scaling behavior due to vacancy-induced DOS has been extensively observed in thermodynamic and magnetization studies on Kitaev QSL materials. We discovered the dynamical scaling law by $\mu$SR. The critical exponent $\alpha = 0.46$ is consistent with $\nu \approx 0.5$, which is theoretically expected for the vacancy-induced low-energy DOS, $N(E) \sim E^{-\nu}$ [23,29,36].

## RESULTS AND DISCUSSION

### Magnetic and thermodynamic properties

The magnetization and heat capacity measurements were performed on powder samples $H_3LiIr_2O_6$. The results are overall consistent with the previous report [23], but discrepancy exists in several details, probably due to different concentration of vacancies. Fig. 2a shows the temperature dependence of dc magnetic susceptibility $\chi(T)$ and its inverse $\chi^{-1}(T)$ of $H_3LiIr_2O_6$, measured under an external field of $\mu_0 H = 1$ T. No trace of magnetic ordering is observed in $\chi(T)$ down to 2 K, in sharp contrast to other Kitaev candidates [16,37-39]. $\chi^{-1}(T)$ shows a linear temperature dependence at high temperatures, indicating a Curie-Weiss behavior. While the linear temperature dependence deviates below 100 K, which may originates from a low-temperature Curie-like contribution due to magnetic defects. The magnetic susceptibility $\chi(T)$ can be well fitted by the function

$$\chi = (1-n)\frac{C}{T-\theta} + n\frac{C_{\mathrm{d}}}{T} + \chi_0 \quad (1)$$

where the two Curie terms are the contribution from the sample and magnetic defects respectively, and $\chi_0$ is the contribution of temperature-independent Van-Vleck susceptibility and core diamagnetic susceptibility. The fitting yields a Curie-Weiss temperature $\theta = -191.6$ K and an effective magnetic moment $\mu = 1.68\,\mu_{\mathrm{B}}$, which are consistent with the previous report and comparable to other honeycomb Ir compounds [23,40]. The concentration of magnetic defects $n$ is estimated to be about 2% yielded from the fitting. It is worth noting that the estimation value of concentration of magnetic defects is not completely equivalent to the concentration of vacancies. The vacancies have various forms including true vacancies and quasi vacancies in $H_3LiIr_2O_6$, which may result in a slightly larger degree of vacancies than 2% actually. Nevertheless, the estimation can still provide a quantitative reference.

The absence of magnetic ordering is further confirmed by the heat capacity measurements down to 0.1 K as shown in Fig. 2b. At zero magnetic field, a low-temperature plateau (with an altitude of 58 mJ $K^{-2}$ mol-$Ir^{-1}$) followed by a divergence in the specific heat coefficient $C/T$ is observed. The divergence at zero field has a temperature dependence of $C/T \sim T^{-0.57}$ as shown in Fig. 2c, almost consistent with $C/T \sim T^{-0.5}$ reported previously [23], and is gradually suppressed by the increasing magnetic fields.

This power-law upturn of $C/T$ is the intrinsic behavior of magnetic specific heat since lattice contribution to the specific heat is negligible below $T = 2$ K. Such a behavior is at odds with the thermodynamics of Kitaev QSL, in which a vanishing specific heat $C/T \sim T^{-0.5}$ is expected [41-43]. Moreover, nuclear Schottky anomaly can be excluded to account for the upturn since it generally survives under magnetic fields of a few Tesla [44], but the low-temperature divergence is significantly suppressed by the applied magnetic fields larger than 1 T. $C/T \sim T^{-0.57}$ indicates that the low energy DOS has a power-law form $N(E) \sim E^{-0.57}$, deviating from the power characteristic of the pure Dirac dispersion of Majorana fermions in Kitaev QSL. Recent theoretical works proposed that a divergent low-energy DOS of $N(E) \sim E^{-\nu}$ with $\nu \approx 0.5$ is expected in the Kitaev model with bond disorder [36], and a finite density of random vacancies in the Kitaev model can also give rise to a pileup of low-energy Majorana eigenmodes, leading to a power-law DOS [23,29,36]. The vacancy effect is not sensitive to the type of vacancies, and can be easily detected even at low concentrations (below 5%) [29], which is consistent with the degree of vacancies (~ 2%) estimated by magnetic susceptibility. The external applied magnetic field plays as a role in breaking time reversal symmetry and removes the divergence of low-energy DOS [29,36].

The scaling plot $B^{1-\alpha_c} C/T$ versus the single dimensionless variable $T/B$ for magnetic fields larger than 1 T is shown in Fig. 2d. The data overlap over 2 orders of magnitude with $\alpha_c = 0.8$, almost consistent with the scaling of $B^{0.5} C/T$ as a function of $T/B$ reported previously for $H_3LiIr_2O_6$ [23]. We note that the low-temperature upturn for data at $\mu_0 H < 3$ T deviates from the data collapse. This is also a characteristic of random-vacancies scenario in the Kitaev QSL where the scaling behavior only applies at relatively high fields [29]. In an attempt to confirm the random-vacancies scenario, we examine the field-temperature scaling of $\chi(T)$ and $M(H)$. The scaling plot $B^{\alpha_m}\chi$ versus $T/B$, and $T^{\alpha_m - 1}M$ versus $B/T$ with $\alpha_m = 0.5$ is shown in Fig. 2d. The scaling behavior of magnetization data generally implies the presence of bond randomness (quasivacancies), which has been reported in other Kitaev materials [40,45-46].

We shall note two differences between our $C/T$ result and the previous report [23]: (i) There is a plateau at low temperatures (0.3 K < $T$ < 1 K) under zero magnetic field in our data, and the upturn exists only below about 0.2 K. The upturn reported before sustains for about 2 orders of magnitude up to 1 K. (ii) The low-temperature upturn is completely suppressed by an external magnetic field larger than 3 T for our sample, whereas $\mu_0 H = 1$ T is enough to suppress the upturn for the sample reported previously. A linear fit of $C \sim \gamma T$ is performed to characterize the plateau in $C/T$ as shown in the inset of Fig. 2b. The theoretical simulation result for Kitaev model with vacancies also shows a plateau-like behavior of $C/T$ before the divergence exists with cooling [29,36], but such a plateau has not attracted much attention and been studied extensively so far. Generally, the plateau in $C/T$ indicates a finite DOS for an insulator. The DOS piles up as temperature decreases, leads to a power-law upturn in $C/T$. Solely contribution from true vacancies cannot account for the upturn of $C/T$ over such a wide range in the previous work [23,29]. However, true vacancies may be dominant in our sample since the power-law upturn of $C/T$ is only observed in the small ultra-low temperature range. The critical value of the magnetic field suppressing the low-temperature upturn probably results from different concentration of vacancies in $H_3LiIr_2O_6$, which still needs more theoretical studies to confirm. Moreover, it is interesting that the low-temperature specific heat at $\mu_0 H = 9$ T is well fitted by $C \sim e^{-\Delta/T}$ with $\Delta = 1.6$ K, exhibiting the characteristic of an energy gap (see the inset of Fig. 2b).

**Magnetic disordered state identified by ZF-$\mu$SR**

The power-law divergence $C/T \sim T^{-0.57}$ at zero magnetic field down to 0.1 K indicates a low energy scale for the power-law DOS. The low-temperature spin dynamics induced by the power-law DOS may exhibit intriguing behavior. $\mu$SR is an ideal technique to detect magnetic ordering and spin dynamics, hence we performed $\mu$SR measurements on $H_3LiIr_2O_6$ to further investigate its magnetic properties. Fig. 3a shows the representative time evolution of the decay position count rate asymmetry $A(t)$, proportional to the positive muon spin polarization $P(t)$ from $T$ = 200 K down to the lowest measured temperature 0.08 K at zero field. A constant background signal ($A_{\mathrm{bg}} = 0.015$) originated from muons that miss the sample and stop in the silver sample holder, has been subtracted from the data. No evidence of magnetic order is found down to 0.08 K, since neither oscillations nor a drastic drop in the initial asymmetry is observed. As the temperature decreases, the initial shape of ZF-$\mu$SR spectra changes from the Gaussian to the Lorentzian like, suggesting that dynamic spin fluctuations may exist at low temperatures.

At zero magnetic field, randomly oriented static (or quasi-static) local internal magnetic field with Gaussian distribution generally gives rise to the Gaussian Kubo-Toyabe depolarization function [47]

$$G_{\mathrm{KT}}(\delta,t)=\frac{1}{3}+\frac{2}{3}\left(1-\delta^2t^2\right)\exp(-\frac{1}{2}\delta^2t^2) \quad (2)$$

where $\delta/\gamma_\mu$ is the distribution width of the local field, and $\gamma_\mu/2\pi=135.53$ MHz/T is the $\mu^+$ gyromagnetic ratio. If local field is dynamic and fluctuates strongly, $P(t)$ is usually well approximated by a Lorentzian exponential function characterized by the muon spin relaxation rate $\lambda$, namely $P(t)\sim\exp(-\lambda t)$.

The asymmetry of ZF-$\mu$SR spectra in $H_3LiIr_2O_6$ cannot be described in terms of only the Kubo-Toyabe type, or the simple exponential function. They are best fitted by the following function

$$A(t)=A_1e^{-\lambda_1 t}G_{\mathrm{KT}}(\delta_1,t)+A_2e^{-\lambda_2 t}+A_3e^{-\lambda_3 t}G_{\mathrm{KT}}(\delta_3,t) \quad (3)$$

where $A_{1,2,3}$ denotes the initial asymmetry for three different depolarization components. The total initial asymmetry $A_1+A_2+A_3$ decreases slightly with cooling. This tiny loss of initial asymmetry is attributed to the time resolution of pulse beamline with pulse width of 80 ns [48,49]. The ratio of $A_1$ to $A_2$ is fixed at the value of 2, determined from LF-$\mu$SR experiments, in which the relaxation is purely due to dynamic fields and the analysis is described in detail later. $A_{1,2,3}$ is nearly temperature independent with $A_1=0.113$, $A_2=0.056$, and $A_3=0.069$. $\delta_{1,3}$ is basically temperature independent, thus it is fixed at an average value $\delta_1=0.34\ \mu\mathrm{s}^{-1}$ and $\delta_3=0.027\ \mu\mathrm{s}^{-1}$. We attribute $\delta_{1,3}$ to the Gaussian field distribution width of nuclear dipolar fields. Fig. 3b shows the temperature dependence of ZF muon spin relaxation rates $\lambda_{1,2,3}$. A drastic increase of $\lambda_{1,2,3}$ is observed below $T$ = 3 K, basically the temperature at which the magnetic specific heat arises. The increase of $\lambda_{1,2,3}$ saturates at about 1 K, and they show a plateau behavior down to 0.08 K. The low-temperature plateau of $\lambda$ generally indicates the persistent spin dynamics and a correlated disordered state [50,51].

**Spin dynamics probed by LF-$\mu$SR**

To further confirm the spin dynamics, we performed LF-$\mu$SR measurements with an external field along the direction of the initial muon spin polarization. The longitudinal-field $\mu_0H=100$ mT was chosen to be much larger than the distribution width of static nuclear dipolar fields $\delta_1/\gamma_\mu=0.4$ mT and $\delta_3/\gamma_\mu=0.03$ mT, which is estimated from the ZF-$\mu$SR results. Thus, the static nuclear dipolar field is completely decoupled [47] and the electronic spin dynamics can be observed clearly. Representative LF-$\mu$SR spectra at different temperatures are shown in Fig. 4a. In order to compare with ZF-$\mu$SR spectra, a constant background signal $A_{\mathrm{bg}}=0.015$ is also subtracted for LF-$\mu$SR spectra. The LF spectra are well described by the function

$$A(t)=A_1e^{-\lambda_1 t}+A_2e^{-\lambda_2 t}+A_3e^{-\lambda_3 t} \quad (4)$$

where $A_{1,2,3}$ has the same meaning with that in Eq. (2), but with slightly smaller values. $\lambda_3$ in Eq. (3) is negligible under magnetic field $\mu_0H=100$ mT, therefore it is fixed to be zero. Temperature dependence of LF muon spin relaxation rates $\lambda_1$ and $\lambda_2$ have the same characteristics, i.e., they gradually increase with cooling and saturate at 0.16 μs$^{-1}$ and 1.3 μs$^{-1}$ respectively below about $T$ = 3 K as shown in Fig. 4b. The relaxation persists at an external LF, $\mu_0 H = 100$ mT is insufficient to decouple the muon spin from internal magnetic field. If the exponential depolarization functions with muon spin relaxation rates $\lambda_1$ and $\lambda_2$ originate from static magnetic field with Lorentzian distribution, then the distribution width of static fields should be $\lambda_1/\gamma_\mu = 0.19$ mT and $\lambda_2/\gamma_\mu = 1.53$ mT. An external LF of 2 mT (15 mT) can fully decouple the relaxation with rate $\lambda_1$ ($\lambda_2$). Thus, $\lambda_{1,2}$ originates from dynamic magnetic fields and the plateau of $\lambda_{1,2}$ below 1 K at ZF is indeed an evidence of persistent spin dynamics in $H_3LiIr_2O_6$.

The temperature independent $A_{1,2,3}$ and the similar temperature dependence of $\lambda_{1,2,3}$ for ZF-$\mu$SR (and $\lambda_{1,2}$ for LF-$\mu$SR) suggest that the three components in Eq. (2) and Eq. (3) are due to distinct $\mu^+$ interstitial stopping sites. In addition, the ratio of $\lambda_2/\lambda_1$ is 5.0(5) for ZF-$\mu$SR measurements, and 5.6(3) for LF-$\mu$SR measurements at low temperatures. Within error bar the same ratio suggests that the two exponential relaxations are due to differences in coupling fields rather than inhomogeneity in the spin dynamics. This is expected when the fluctuation rates at the different sites are the same and the correlation length for the fluctuations is short [52,53], and therefore is additional evidence for three muon stopping sites. Since there are various vacancies and disorders in the crystal structure of $H_3LiIr_2O_6$, it is unable to determine the muon stopping sites only by calculation. The nuclear moments of Li and H ions are significant, which can account for the Kubo-Toyabe relaxation in the first term. The Kubo-Toyabe relaxation is absent in the second term, and $\delta_3$ in the third term is so small, an order of magnitude smaller than $\delta_1$. Thus, the second and third term possibly arise from the muons, whose stopping sites are not near H ions due to random substitution of $H^+$.

The low-temperature spin dynamics in $H_3LiIr_2O_6$ revealed by $\mu$SR shares the same temperature range with the growing magnetic specific heat, hence very likely also originates from the vacancy-induced low-energy DOS. The divergent low-energy DOS of $N(E) \sim E^{-0.57}$ obtained from specific heat results may lead to time-field scaling relations in muon polarization. LF-$\mu$SR measurements under various LF up to 0.3 T at $T$ = 0.1 K and 4 K were carried out to check the scaling relation and investigate the spin dynamic autocorrelation function. The low-energy DOS under magnetic field up to 0.3 T can be regarded unchanged compared to the ZF case. Because the low-temperature upturn of $C/T$ at 1 T can also be characterized by the power-law with the power -0.61, which is very closed to -0.57 of ZF case (see Fig. 2c). The field dependence of $\lambda_1$ and $\lambda_2$ is shown in Fig. 5a. $\lambda_1$ ($\lambda_2$) at 0.1 K and $\lambda_2$ at 4 K at $B$ > 5 mT show consistent power-law field dependence with power close to -0.46. We note the field dependence of $\lambda_1$ at 4 K is not used to characterize the power law, because $\lambda_1$ at 4 K is so small (of the order of 0.01 μs$^{-1}$), it is much more reasonable and accurate to get the power from $\lambda_2$. The power-law behavior of $\lambda(B)$ indicates the time-field scaling of LF-$\mu$SR spectra [54,55]. Therefore, we plot the time spectra as a function of the scaling variable $t/B^{\alpha}$ with $\alpha = 0.46$ in Fig. 5b. The observation of the time-field scaling can be interpreted as a signature of cut-off power-law spin autocorrelation function, $q(t) = \langle S_i(t) \cdot S_i(0) \rangle \sim t^{\alpha-1} \exp(-\lambda t)$ for relatively high fields [54,55].

The scaling behavior of LF-$\mu$SR is not the only signature for spin dynamics in Kitaev materials. It was widely observed in many spin-glass materials above the freezing temperature as a result of slow spin dynamics [54,56]. For the spin glass, the asymmetry is generally fitted with the stretched exponential function. The stretched factor decreases to about 1/3 when the system is approaching the freezing temperature [52], and the scaling does not apply anymore. However, no spin freezing is observed, and the scaling law keeps valid down to 0.1 K in $H_3LiIr_2O_6$. For the pure Kitaev model, the spin operator is written in terms of Majorana fermions $S_i^a = ib_i^a c_i/2$, where $b_i^a$ is related to the flux and $c_i$ represents the low-energy Majorana excitations. The localized Majorana fermion $b_i^a$ carries the gauge flux and is gapped in the Kitaev spin liquid, behaving like typical gapped particles. The time evolution is generally approximated by $\exp(-\lambda t)$, where $\lambda$ represents the lifetime of the gauge flux. If the lifetime $\lambda$ is very long, the time evolution can be approximated as a constant. In the case of $\mu$SR, the stretched exponential time evolution function $\exp[(-\lambda t)^{\beta}]$ is also commonly used. Regardless of whether the time evolution function is constant, exponential, or stretched exponential, the scaling behavior remains valid [55]. The spin autocorrelation function is $q(t) = \langle S_i(t) \cdot S_i(0) \rangle \sim \langle c_i(t) \cdot c_i(0) \rangle \exp(-\lambda t)$ when the time evolution is dominated by the low-energy excitations $c_i$ and the time revolution of $b_i^a$ gives rise to $\exp(-\lambda t)$. Thus, $q(t)$ is related to the Green function of the low-energy Majorana excitation $c_i$, and can be taken as the Fourier transformation of the DOS

$$\begin{aligned} q(t) &\sim \exp(-\lambda t) \int N(E) \exp(-iEt) \mathrm{d}E \\ &= t^{\alpha-1} \exp(-\lambda t) \int Et^{-\alpha} \exp(-iEt) \mathrm{d}(Et) \\ &\sim t^{\alpha-1} \exp(-\lambda t) \end{aligned} \tag{5}$$

Therefore, the time-field scaling of LF-$\mu$SR polarization $A(t) \sim t/B^{\alpha}$ ($\alpha = 0.46$) in Fig. 5b implies that the low-energy DOS has a power-law form $N(E) \sim E^{-\alpha}$ with $\alpha = 0.46$. This critical exponent is close to $\alpha = 0.57$ obtained from low-temperature specific heat results, and they are both consistent with $N(E) \sim E^{-\nu}$ ($\nu \approx 0.5$), which is expected for random-vacancies scenario in Kitaev QSL [29].

To conclude, our thermodynamic and ZF-$\mu$SR measurements reveal no sign of magnetic ordering or glass transition down to 0.08 K, confirming a disordered correlated ground state in $H_3LiIr_2O_6$. The low-temperature power-law specific heat coefficient $C/T \sim T^{-0.57}$ signals the finite DOS with the form $N(E) \sim E^{-0.57}$, in a good agreement with the random-vacancies scenario for Kitaev QSL model. The scaling laws observed in specific heat and magnetization data further support this scenario. LF-$\mu$SR measurements were carried out to investigate the spin dynamics associated with vacancy-induced DOS. We found not only dynamical spin fluctuation persists at low temperatures, but also LF-$\mu$SR polarization intriguingly shows a time-field scaling behavior $A(t) \sim t/B^{0.46}$, which implies a power-law spin autocorrelation function consistent with the divergent DOS.

Many theoretical works have been proposed to explain the peculiar features reported for the Kitaev QSL candidate $H_3LiIr_2O_6$. Motivated by the random-vacancies scenario, we revisited the specific heat and magnetic properties of $H_3LiIr_2O_6$ and performed $\mu$SR measurements. The scaling laws of specific heat, magnetization and LF-$\mu$SR data show good consistency with the critical exponent close to theoretical value $\nu \approx 0.5$, indicating the same physics of DOS $N(E) \sim E^{-0.5}$. The time-field scaling behavior of LF-$\mu$SR polarization provides a crucial way for determining dynamical properties of vacancy-induced DOS, while scaling laws have so far been focused on static aspects in thermodynamics and magnetization studies. Our work provides more experimental evidence for the divergent DOS in a Kitaev QSL with vacancies, particularly presents $\mu$SR results to understand the low-energy spin dynamics, which offers a new insight into the nature of vacancies in $H_3LiIr_2O_6$.

## METHODS

### Sample preparation

Polycrystalline samples of $H_3LiIr_2O_6$ were prepared by the cation exchange reaction [27,28]. The starting materials $Li_2CO_3$ and $IrO_2$ powder were mixed thoroughly in the ratio of 1.05: 1, and placed in an alumina crucible with a lid and annealed at 840 °C for one day to obtain the precursor $\alpha$-$Li_2IrO_3$ powder. For the cation exchange, $\alpha$-$Li_2IrO_3$ powder was added into a Teflon-lined steel autoclave with 4 mol/L $H_2SO_4$ aqueous solution. The mixture was heated in the sealed vessel at 120 °C for one hour. Polycrystalline $H_3LiIr_2O_6$ was obtained by washing with distilled water and drying at 80 °C.

### Sample characterization

The PXRD measurements were conducted on Rigaku Smartlab 9KW using Cu K$\alpha$ radiation at room temperature. The PXRD pattern of $H_3LiIr_2O_6$ in Fig. 1b shows that several diffraction pattern peaks are broaden and even disappear, suggesting the heavily stacking faulted crystal structure. Our PXRD pattern agrees with the previous results [22,23], and the stable phase $H_3LiIr_2O_6$ was confirmed from the synthetics evolution process ($\alpha$-$Li_2IrO_3$→$H_3LiIr_2O_6$→ $H_5LiIr_2O_6$) [24].

**Magnetic measurements**

Magnetic susceptibility and isothermal magnetization measurements were carried out on a Quantum Design SQUID magnetometer down to 1.8 K.

**Thermodynamic measurements**

The specific heat was measured on a Quantum Design Physical Properties Measurement System (PPMS) equipped with Dilution Refrigerator using heat-pulse technique.

***μ*SR measurements**

$\mu$SR experiments were carried out on MuSR spectrometer at ISIS Neutron and Muon Facility, Rutherford Appleton Laboratory Chilton, UK (temperature down to 80 mK, longitudinal external magnetic field up to 0.3 T). Spin-polarized positive muons were implanted into 1 gram of polycrystalline sample mounted on a silver sample holder covering a circular area of 1 inch in diameter.

## DATA AVAILABILITY

The data that support the findings of this study are available from the corresponding author upon reasonable request.

## ACKNOWLEDGMENTS

We are grateful to the ISIS cryogenics Group for their valuable help during the $\mu$SR experiments (10.5286/ISIS.E.RB1910121). The work at Fudan University was supported by the National Natural Science Foundation of China under Grant No. 12034004 and No. 12174065, and the Shanghai Municipal Science and Technology, Major Project Grant No. 2019SHZDZX01. The work at SUSTech was supported by Shenzhen Fundamental Research Program (Grant No. JCYJ20220818100405013 and JCYJ20230807093204010).

## AUTHOR CONTRIBUTIONS

L. S. and J. M. conceived the project. L. H., C. L., L. W., Y. S. and F. Y. synthesized and characterized the sample. Y. Y. performed the thermodynamic and magnetization measurements. Y. Y., Z. Z., C. C., Q. W., Z. D., and C. T. carried out the $\mu$SR experiments. A. D. H. and P. K. B provided support and expertise at beam lines during $\mu$SR experiments. Y. Y. and C. J. analyzed the data. K. C. provided support on analyzing the muon stopping sites. Y. Y., C. J., J. M. and L. S. wrote this manuscript with input from all authors. Y. Y., C. J., and L. H. contributed equally to this work.

## COMPETING INTERESTS

The authors declare no competing financial or non-financial interests.

## ADDITIONAL INFORMATION

**Correspondence** should be addressed to Prof. Jia-Wei Mei (meijw@sustech.edu.cn), and Prof. Lei Shu (leishu@fudan.edu.cn).

**FIGURES AND CAPTIONS**

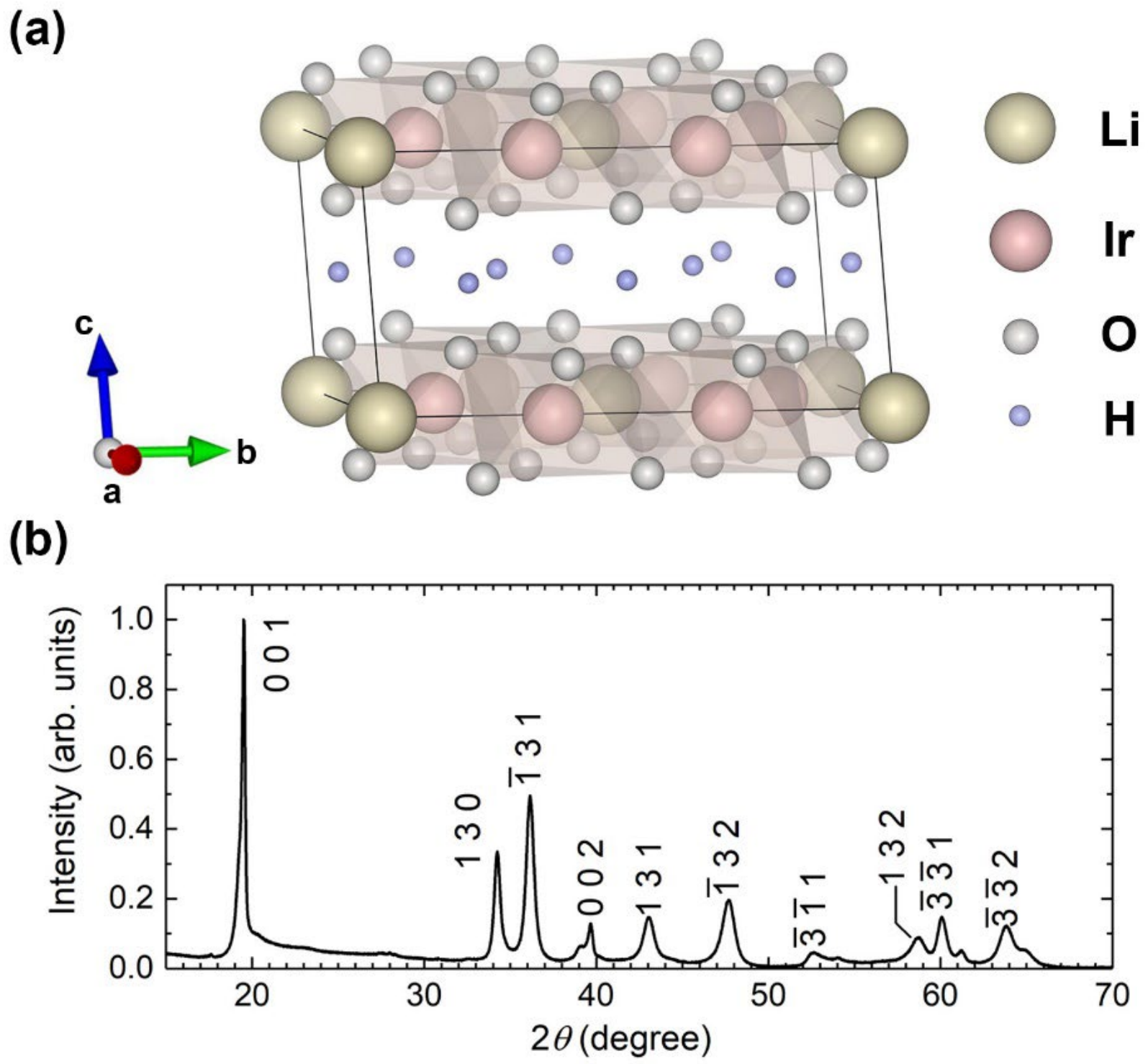

**Fig. 1 Sample characterization and crystal structure of $H_3LiIr_2O_6$.** (a) Crystal structure of $H_3LiIr_2O_6$. Yellow spheres: $Li^+$ ions. Pink spheres: $Ir^{4+}$ ions. White spheres: $O^{2-}$ ions. Purple spheres: $H^+$ ions. The $Ir^{4+}$ ions form honeycomb lattice in *ab* plane. (b) Powder XRD pattern of $H_3LiIr_2O_6$ at room temperature.

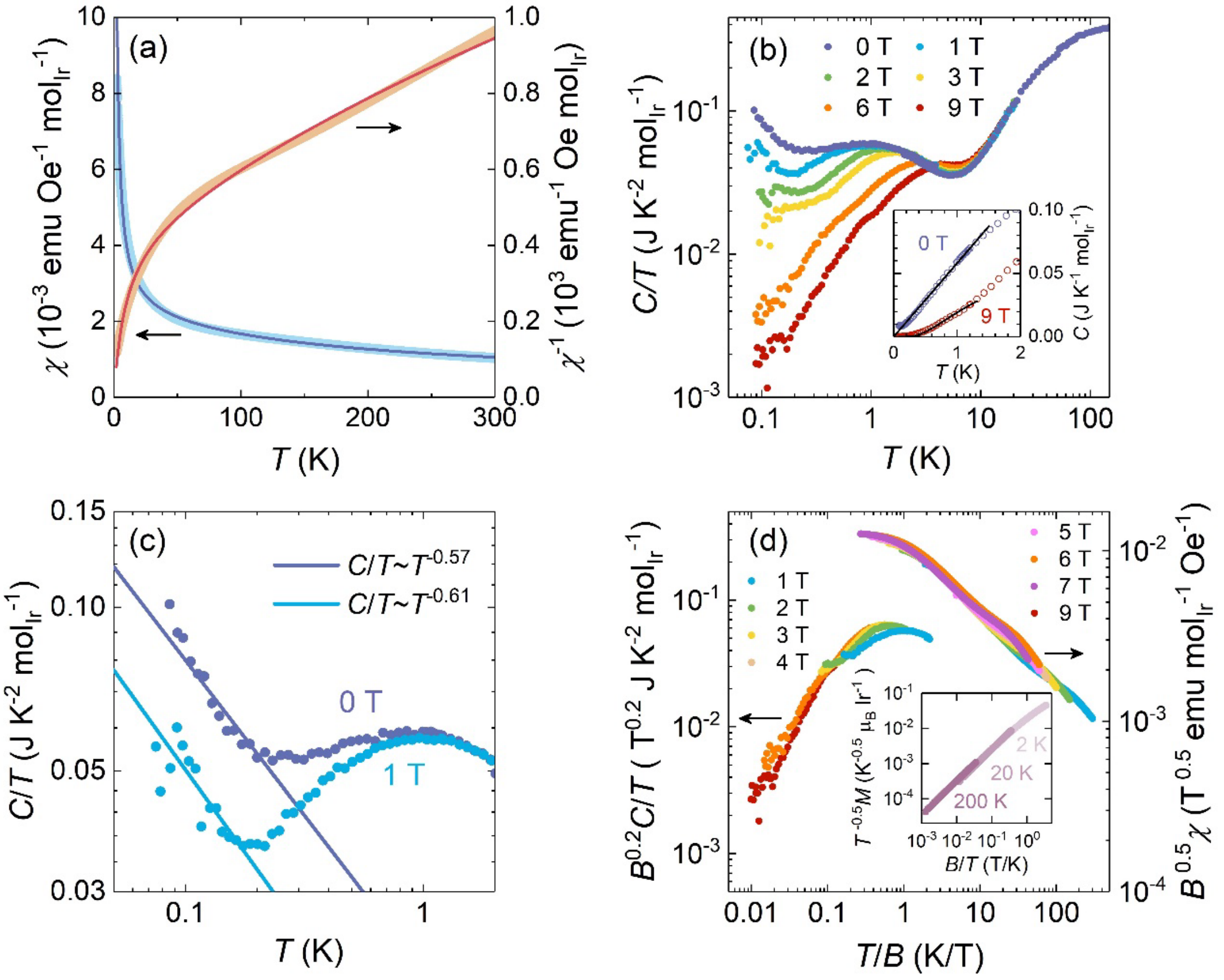

**Fig. 2 Magnetic susceptibility and specific heat of $H_3LiIr_2O_6$.** (a) Temperature dependence of magnetic susceptibility $\chi$ and its inverse $\chi^{-1}$ of $H_3LiIr_2O_6$ measured at $\mu_0 H = 1$ T. The blue and red solid lines are the fit and the inverse of fit using Eq. (1) respectively. (b) The specific heat coefficient $C/T$ as a function of $T$ under various magnetic fields. Inset: Temperature dependence of $C$ at 0 T and 9 T. The solid curves are low-temperature fitting $C \sim \gamma T$ for 0 T and $C \sim e^{-\Delta/T}$ for 9 T. (c) The specific heat coefficient $C/T$ as a function of $T$ below 2 K at zero magnetic field and 1 T respectively. The total specific heat can be regarded as magnetic specific heat since the lattice contribution is negligible below 2 K. The solid purple and blue lines represent power-law fitting for the low-temperature upturns. (d) Data collapses for $B^{0.2}C/T$ vs $T/B$, and $B^{0.5}\chi$ vs $T/B$. Inset: Data collapse for $T^{-0.5}M$ vs $B/T$.

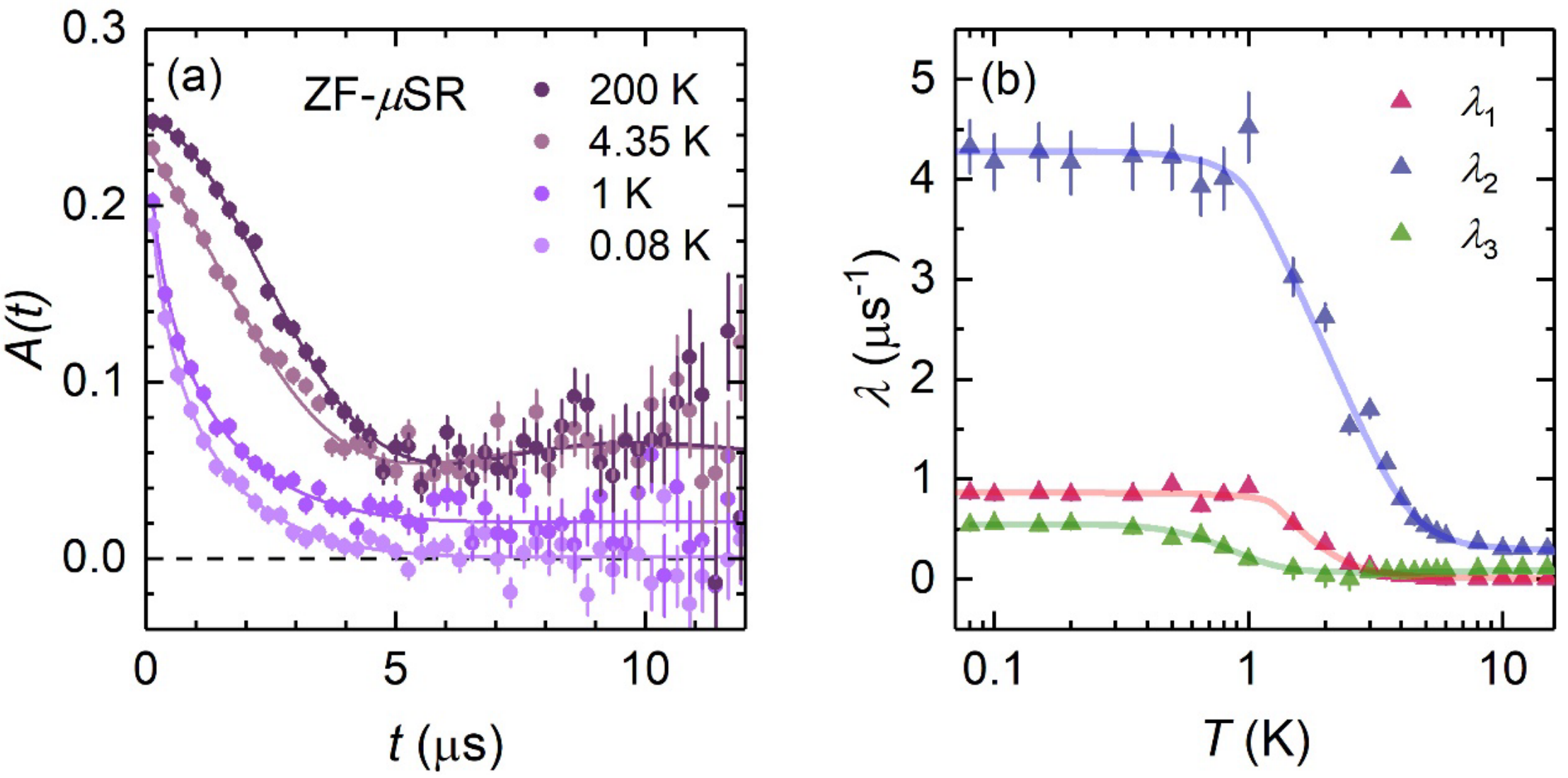

**Fig. 3 ZF-$\mu$SR results for $H_3LiIr_2O_6$.** (a) ZF-$\mu$SR spectra of $H_3LiIr_2O_6$ at selected temperatures. Colored lines are the fits to raw data using Eq. (2). (b) Temperature dependence of muon spin relaxation rates $\lambda_{1,2,3}$. The solid lines are a guide to the eye. $\lambda_{1,2,3}$ shows similar a plateau behavior at low temperatures.

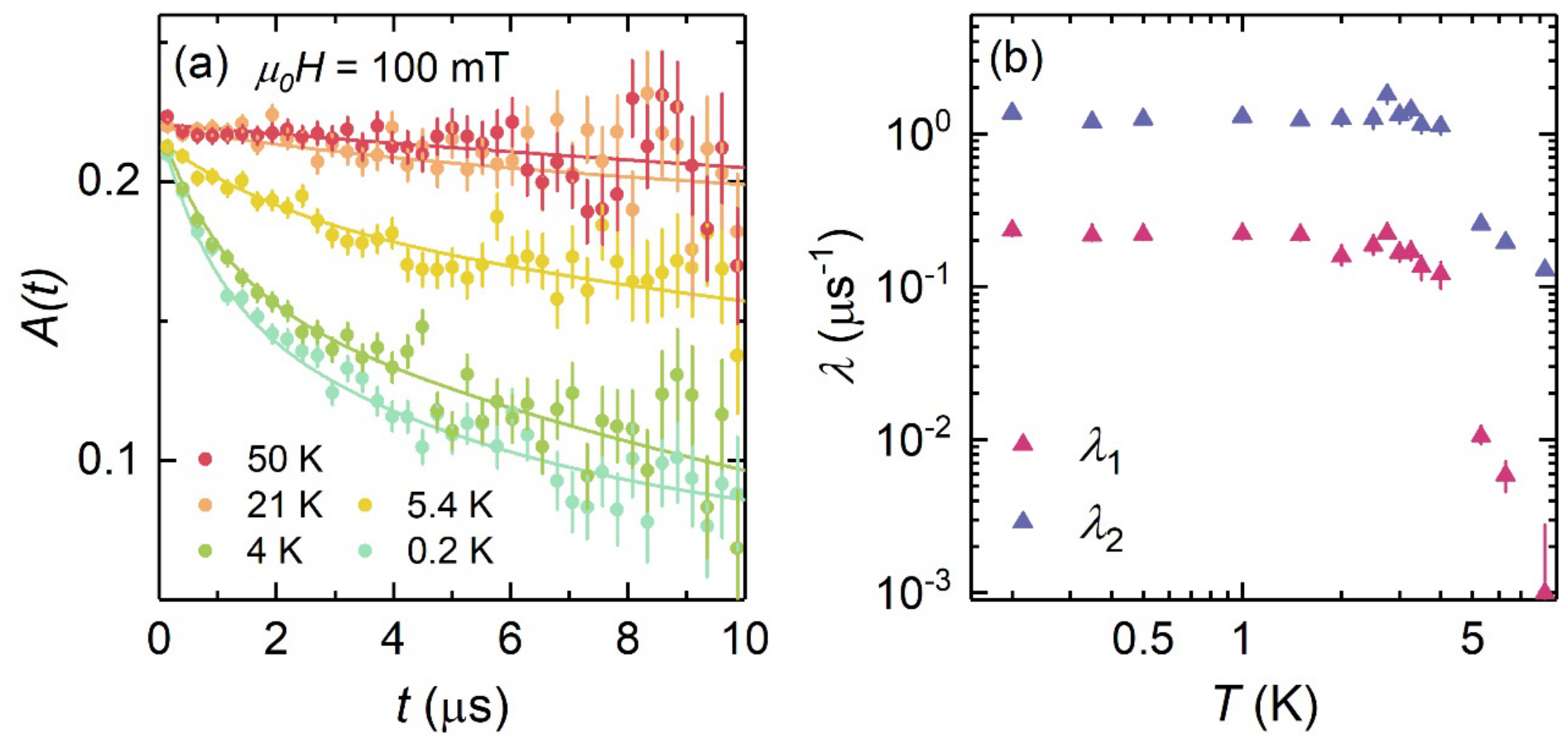

**Fig. 4 LF-$\mu$SR results for $H_3LiIr_2O_6$ under 100 mT.** (a) LF-$\mu$SR spectra of $H_3LiIr_2O_6$ under $\mu_0 H = 100$ mT at selected temperatures. Colored lines are fits to the raw data using Eq. (3). The relaxation persists at low temperatures. (b) Temperature dependence of LF muon spin relaxation rates $\lambda_{1,2}$. They both show a plateau below about $T = 3$ K.

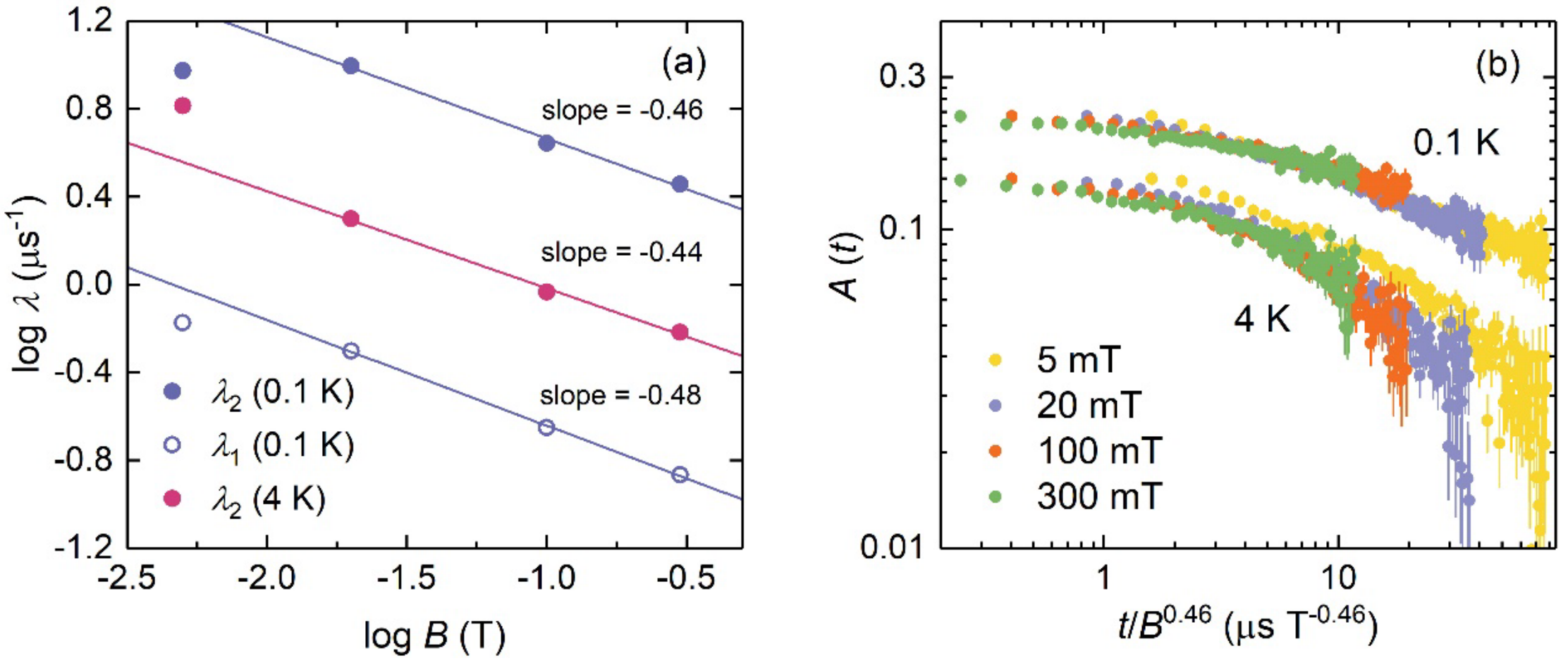

**Fig. 5 Time-field scaling behavior of LF-$\mu$SR polarization.** (a) The logarithm of LF muon spin relaxation rates $\lambda_{1,2}$ as a function of the logarithm of applied magnetic fields. The solid lines are a linear fit for the data at $B$ > 5 mT. (b) Time-field scaling of LF-$\mu$SR spectra at 0.1 K and 4 K for $H_3LiIr_2O_6$.